\documentclass[aps, pre, eps, superscriptaddress, twocolumn, amsmath,amssymb,floatfix]{revtex4}
\newif\ifpdf\ifx\pdfoutput\undefined\pdffalse\else\pdfoutput=1\pdftrue\fi

\usepackage{graphicx}
\usepackage{subfigure}
\usepackage{bm}
\usepackage{epsfig}

\newcommand{\musig}{\mu(\sigma)}
\newcommand{\p}{^{(0)}}
\newcommand{\one}{^{(1)}}
\newcommand{\two}{^{(2)}}
\renewcommand{\a}{^{(a)}}
\newcommand{\ronesig}{\rho\one(\sigma)}
\newcommand{\rtwosig}{\rho\two(\sigma)}
\newcommand{\rbarsig}{\bar\rho(\sigma)}
\newcommand{\parent}{\rho\p(\sigma)}
\newcommand{\f}{f}
\newcommand{\fparent}{\f\p(\sigma)}
\newcommand{\cl}{_{\rm cl}}

\newcommand{\bmu}{{\boldsymbol{\mu}}}
\newcommand{\brho}{\boldsymbol{\rho}}
\newcommand{\bchi}{\boldsymbol{\chi}}
\newcommand{\bzero}{\boldsymbol{0}}

\newcommand{\bff}{\mathbf{\f}}

\newcommand{\DP}{\Delta P}

\newcommand{\be}{\begin{equation}}
\newcommand{\ee}{\end{equation}}
\newcommand{\bea}{\begin{eqnarray}}
\newcommand{\eea}{\end{eqnarray}}

\begin{document}

\title{\bf Accurate simulation estimates of cloud points of polydisperse fluids}

\author{Matteo Buzzacchi}
\affiliation{Department of Physics, University of Bath, Bath BA2 7AY,
United Kingdom}
\author{Peter Sollich}
\affiliation{King's College London, Department of Mathematics, Strand, London WC2R 2LS, United Kingdom}
\author{Nigel B. Wilding}
\affiliation{Department of Physics, University of Bath, Bath BA2 7AY, United Kingdom}
\author{Marcus M\"uller}
\affiliation{Institut f\"ur Theoretische Physik,
Georg-August-Universit\"{a}t, Friedrich-Hund Platz 1, D-37077 G\"ottingen, Germany}
\date{\today}

\begin{abstract} 

We describe two distinct approaches to obtaining cloud point densities
and coexistence properties of polydisperse fluid mixtures by Monte Carlo
simulation within the grand canonical ensemble. The first method
determines the chemical potential distribution $\mu(\sigma)$ (with
$\sigma$ the polydisperse attribute) under the constraint that the
ensemble average of the particle density distribution
$\rho(\sigma)$ matches a prescribed parent form. Within the region of
phase coexistence (delineated by the cloud curve) this leads to a
distribution of the fluctuating overall particle density $n$, 
$p(n)$, that necessarily has unequal
peak weights in order to satisfy a generalized lever rule. A theoretical
analysis shows that as a consequence, finite-size corrections to
estimates of coexistence properties are {\em power laws} in the system
size. The second method assigns $\mu(\sigma)$ such that an {\em equal
peak weight} criterion is satisfied for $p(n)$ for all points within the
coexistence region. However, since equal volumes of the coexisting
phases cannot satisfy the lever rule for the prescribed parent, their
relative contributions must be weighted appropriately when determining
$\mu(\sigma)$. We show how to ascertain the requisite weight factor
operationally. A theoretical analysis of the second method suggests that
it leads to finite-size corrections to estimates of coexistence
properties which are {\em exponentially small} in the system size. The
scaling predictions for both methods are tested via Monte Carlo
simulations of a novel polydisperse lattice gas model near its cloud
curve, the results showing excellent quantitative agreement with the
theory.

\end{abstract} 
\maketitle
\setcounter{totalnumber}{10}

\section{Introduction and background}

Examples of polydisperse fluids arise throughout soft matter science,
notably in colloidal dispersions, polymer solutions and liquid-crystals.
Typically the polydispersity of such systems is manifest as variation in
some attribute such as particle size, shape or charge, which is
customarily denoted by a continuous parameter $\sigma$. The state of the
system is then quantifiable in terms of a distribution $\rho(\sigma)$
measuring the number density of particles of each $\sigma$; more
precisely, $\rho(\sigma)d\sigma$ is the number density of particles in
the range $\sigma\ldots\sigma+d\sigma$. As such, 
one can regard the system as a mixture of an infinite number of particle
``species'' each labelled by the value of $\sigma$ \cite{SALACUSE}.

As has long been appreciated, polydispersity can profoundly influence
the thermodynamical and processing properties of complex fluids
\cite{LARSON99,CHAIKIN00}, making a clear elucidation of its detailed
role a matter of both fundamental and practical importance. In
particular the  phase behaviour of polydisperse systems is known to be
considerably richer in both variety and character than that of
corresponding monodisperse systems (see \cite{e_g_my_review} for a
recent review). The source of this richness can be traced to {\em
fractionation} effects: at coexistence a polydisperse fluid described by
some initial ``parent'' distribution, $\rho^{(0)}(\sigma)$, may divide
into two or more ``daughter'' phases $\rho^{(a)}(\sigma)$, $a=1,2,...$,
each of which differs in composition from the parent. The sole
constraint is that the volumetric average of the daughter distributions
equals the parent distribution.  

\begin{figure}
\includegraphics[width=7.0cm,clip=true]{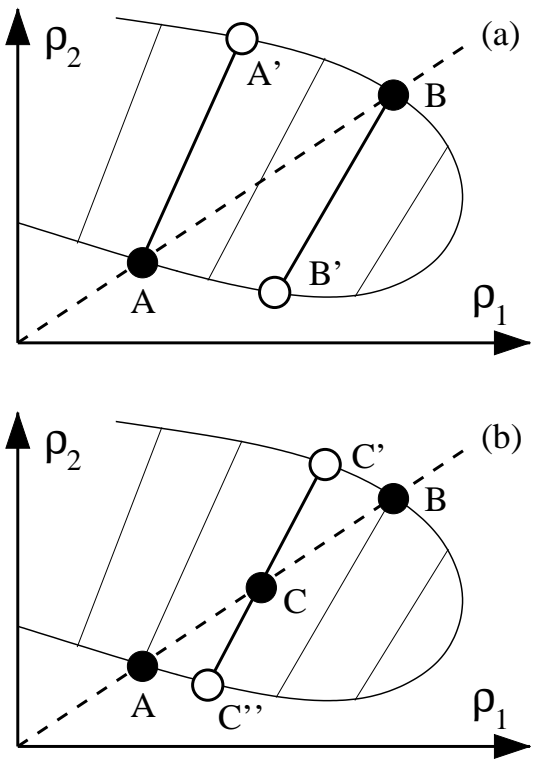}
\caption{Schematic representation of the fractionation behaviour of a binary
fluid mixture, as described in the text. The parent densities
$\rho\p_1,\rho\p_2$ are constrained to the dilution line (dashed
curve). The curved line indicates the boundary of the coexistence region;
straight tielines connect coexisting phases. {\bf (a)} At the cloud 
points A and B, the parent phase coexists with the shadow phase
A' and B' respectively. {\bf (b)} For a general parent
density C in the coexistence region, two 
daughter phases C' and C'' form. The parent
lies on the tieline connecting them; this is the geometrical
representation of the lever rule.}
\label{fig:schem}
\end{figure}

The occurrence of fractionation can engender dramatic alterations to
phase diagrams. Insight into the essential features of
polydisperse phase behaviour can be gained by first considering the
simpler case of a binary mixture of two components whose densities we
denote $\rho_1$ and $\rho_2$ (see also Ref.~\cite{e_g_my_review}).  Let
us confine our attention to the case of a ``dilution line'' in the full
phase diagram, in which we vary (at some fixed temperature) the overall
(parent) density $n\p=\rho_1^{(0)}+\rho_2^{(0)}$, whilst holding
constant the {\em ratio} of the densities $\rho_1^{(0)}/\rho_2^{(0)}$,
i.e.\ the overall composition. These parents thus lie on a straight line
through the origin in the $(\rho_1,\rho_2)$-plane, as shown in
fig.~\ref{fig:schem}(a). So-called cloud points (marked A and B) delimit
the range of parent densities for which phase coexistence occurs on the
dilution line. At cloud point A, the low density parent phase coexists
with an infinitesimal volume of a high density daughter phase A'; while
at cloud point B, the parent coexists with an infinitesimal volume of a
low density daughter phase B'. Owing to fractionation,  however, the
compositions $\rho_1\a/\rho_2\a$ $(a=1,2)$ of the incipient daughter or
``shadow'' phases differ in general from that of the parent: the shadow
points (A' and B') lie {\em off} the dilution line.

For parents on the dilution line but with densities $n\p$ intermediate
between the cloud points (e.g.\ point C of fig.~\ref{fig:schem}(b)), two
daughter phases (C' and C'') form. These phases
occupy finite fractional volumes and their compositions
$\rho\a_1/\rho\a_2$ $(a=1,2)$ {\em both} differ
from that of the parent. Moreover, the daughter phase compositions vary
non-trivially as one scans the parent density $n\p$ between the
cloud points. Consequently, and as a result of the lever rule
$(1-\xi)\rho\one_i +\xi\rho\two_i=\rho\p_i$ ($i=1,2$), the
fractional volumes $1-\xi$ and $\xi$ of two phases will in general depend non-linearly on the
parent density. To fully specify the coexistence properties, one thus
needs to determine the variation of $\xi$ and
$\rho\a_1/\rho\a_2$ $(a=1,2)$ with  $n\p$. 

\begin{figure}
\includegraphics[width=7.0cm]{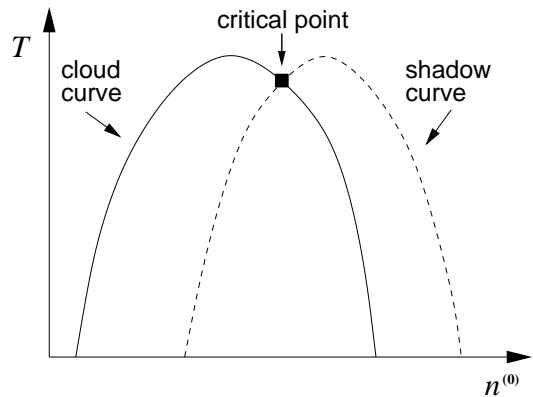}
\caption{A schematic fluid-fluid phase diagram for a polydisperse
fluid, showing temperature $T$ against density; the cloud curve gives
the parent density $n\p$ where phase separation first occurs while the
shadow curve records the density of the incipient coexisting phases.}
\label{fig:phasediagram}
\end{figure}

Turning now to the fully polydisperse case, we consider a family of
parent density distributions $\parent=n\p\fparent$ with fixed
(normalized) particle size distribution $\fparent$ and varying overall
number density $n\p$, the value of which parameterizes the location of
the system on the dilution line. Repeating the 
above considerations for a range of temperatures, one finds that
the familiar liquid-vapor binodal in the density-temperature plane of a
monodisperse fluid splits into cloud and shadow curves
\cite{e_g_my_review}, as shown schematically in
fig.~\ref{fig:phasediagram}. These mark, respectively, the density of
the onset of phase separation and the density of the incipient (shadow)
phase \cite{NOTE0}. The critical point no longer
occurs at the maximum of the cloud curve but at the intersection of the
cloud and shadow curves (at which both coexisting phases are identical).
One interesting implication of this is that even at the critical temperature, liquid
vapor coexistence can occur provided the overall parent density is less
than its critical value. 

In this work we address the problem of accurately determining phase
coexistence properties or, more specifically, cloud point densities of
polydisperse mixtures via Monte Carlo (MC) simulation. In a monodisperse
system, this task is relatively straightforward (see e.g.\ 
Ref.~\cite{WILDING95}), because the properties of the coexisting phases
at a given temperature are the same for all values of the overall
density lying within the coexistence region (as delineated by the
binodal). By contrast, for multicomponent fluids, the occurrence of
fractionation implies (as we have seen) that the compositions of the
phases {\em vary} across the coexistence region (i.e.\ as a function of
parent density $n\p$) and one needs to consider carefully the
repercussions of this for simulation estimates of coexistence
properties. 

The layout of our paper is as follows. In the following section we
describe the principal computational issues arising from fractionation
and detail two distinct strategies for locating cloud point densities
within a grand canonical ensemble framework. The finite-size scaling
properties of both methods are analyzed theoretically by generalizing to
polydisperse systems scaling concepts developed originally in the
context of monodisperse phase equilibria. The predictions for both
methods are tested in sec.~\ref{sec:simulations} via detailed MC
simulations of a novel polydisperse lattice gas model. We conclude in
sec.~\ref{sec:concs} by comparing and contrasting the relative merits of
the two approaches.

\section{Methods for cloud point determination}
\label{sec:methods}


We shall work within the grand canonical ensemble (GCE), where the set
of chemical potentials $\musig$ is the control parameter, while the
particle density $n$ and the particle size distribution are fluctuating
variables. (Temperature is assumed held fixed and not written
explicitly.) The existence of two phases at given chemical potentials
can be detected from the presence of two separate peaks in the
probability distribution of the fluctuating order parameter, which we
take to be the number density $n$. The fractions $x\one$, $x\two$ of
probability mass under each of the two peaks of $p(n)$ (i.e.\ the peak
weights) are related to the pressure difference between the two phases
as discussed below. For an infinitely large system, where the GCE and
the canonical ensemble become equivalent, $x\one$ and $x\two$ would
correspond to the fractions of overall system volume occupied by each
phase, i.e.\ $1-\xi$ and $\xi$ as defined above.

The density distributions $\ronesig$, $\rtwosig$ of the two phases can
be assigned by averaging only over configurations belonging to either
peak of $p(n)$. These are related to the overall average density distribution by

\be 
\rbarsig = x\one \ronesig + x\two \rtwosig\:, 
\label{rhobar} 
\ee
which can be regarded as a generalized form of the lever rule.

In order to estimate the cloud point for the set of dilution line parents
$\parent=n\p\fparent$, one needs to find the value $n\p\cl$ of the
parent density $n\p$ at the boundary of the coexistence region, where
the fractional volume of one of the phases, say $x\two$, drops to zero.
Of course, a value of exactly zero is obtained only in the thermodynamic
limit of infinite system size; the key question is how a reliable
estimate of $n\p\cl$ can nevertheless be extracted from data for
finite-sized systems.

\subsection{Method A}
\label{sec:methodA}

In method A, proposed originally in~\cite{WilSolFas05}, we take the
natural step of tuning the chemical potentials $\musig$ such that
$\rbarsig$ from (\ref{rhobar}) equals directly the parent $\parent$, and
measure the density-dependence of the peak weight ratio $r=x\two/x\one$.
(We do not detail here the algorithm for tuning the $\musig$, which is
explained in~\cite{musig_tuning}). To understand how $r$ will depend on
parent density $n\p$ and system size, note that $r$ is determined by the
difference in the grand potential; this is directly related to the
pressure $P$ so that $r=\exp(L^d\beta \DP)$ for large system size $L$.
Here $d$ is the dimension of space, $\beta=1/k_{\rm B}T$ and $\DP =
P\two - P\one$. The criterion for stable coexistence is that $r$ must
have a finite value as $L\to\infty$; the pressure difference then has to
scale as $\DP \sim L^{-d}$ except in the special case $r=1$ (see method
B below).

For finite $L$, metastable coexistence can still be observed in the
density region $n\p<n\p\cl$ where $\Delta P=O(1)$, but here $r$ will
be exponentially small. To estimate $n\p\cl$ from data for a finite
system, we use the fact that $\DP$ is $O(1)$ and scales linearly with
$n\p-n\p\cl$ to leading order near the cloud point, and hence $\ln
r\sim L^d(n\p-n\p\cl)$. This applies for $n\p<n\p\cl$, while above
$n\p\cl$ one has $\ln r=O(1)$. Thus the derivative $(\partial/\partial
n\p)\ln r$ should drop from an $O(L^d)$ plateau to $O(1)$ around
$n\p=n\p\cl$. In the second derivative $-(\partial/\partial n\p)^2\ln
r$ this drop will manifest itself as a peak, whose position serves as an
estimate for $n\p\cl$. We next derive the finite-size scaling of the
location and shape of this peak.

Our starting point is the rigorous result of~\cite{BorKot92} for
first-order phase transitions driven by some field $h$. The authors
of~\cite{BorKot92} showed that, if the transition is at $h=0$, the
free energy density around this point is given by
\be
f(h)=-L^{-d}k_{\rm B}T\ln\left(e^{-L^d \beta f\one(h)} + e^{-L^d \beta
f\two(h)}\right)
\label{BorKot}
\ee
up to exponentially small corrections of order $L^{-d}
\exp(-\mbox{const}\times L)$, which can be neglected. The function
$f\one(h)$ ($f\two(h)$) is the thermodynamic free energy of the first
(second) phase in the regime where that phase is stable; elsewhere,
i.e.\ for values of $h$ where the phase would be only metastable in an
infinite system, it can be
chosen as the continuation with smooth derivatives (up to at least 3rd
order~\cite{BorKot92}) of the stable free energy. Intuitively, the
approximation (\ref{BorKot}) tells us that the partition function
close to the phase transition can be obtained by adding the
partition functions of the coexisting (stable or metastable) phases.
Formally, it is valid only for values of $|h|\ll L^{-1}$; but in fact
we will be interested in rather smaller values of $h\sim L^{-d}\ln
L$ where the smaller phase is not yet exponentially suppressed, so
that this is not a restriction.

For a polydisperse system within the GCE, the analogue of the field
$h$ is the set of chemical potentials. Conceptually it is easiest to
think first of particle sizes discretized into a large but finite
number $M$ of bins; there are then $M$ chemical potentials and
densities which we write as vectors $\bmu$ and $\brho$,
respectively. More precisely, we take $\bmu$ as the difference of the
chemical potentials from those at the desired cloud point, so that the
latter is located at $\bmu=\bzero$. Assuming that the result
(\ref{BorKot}) can be extended to situations with $M$ field variables
instead of a single one we have then for the negative grand potential density,
which is nothing but the pressure,
\be
P(\bmu)=L^{-d}k_{\rm B}T\ln
\left(e^{L^d \beta P\one(\bmu)} + e^{L^d \beta P\two(\bmu)}\right)
\label{BorKot_poly}
\ee
The single-phase pressures expanded to second order in $\bmu$ read
($a=1,2$)
\be
P\a(\bmu) = P\cl + \brho\a\cl \cdot \bmu + \frac{1}{2}\bmu\cdot
\bchi\a\bmu
\label{quadr_expansion}
\ee
%
Here $P\cl$ is the coexistence pressure at the cloud point, which is
common to both phases, while $\brho\one\cl$ and $\brho\two\cl$ are the
(vectors of) particle densities at the cloud point; $\brho\one\cl$ is
then equal to the parent density vector $\brho\p\cl$ at the cloud point,
while $\brho\two\cl$ is the shadow density vector. The matrices
$\bchi\a=\nabla_\bmu \brho\a=\nabla_\bmu \nabla_\bmu P\a$ are the
susceptibilities of the particle densities to chemical potential
changes, again evaluated at the cloud point.

Taking the chemical potential derivative of (\ref{BorKot_poly}) we
have for the overall density vector
\be
\bar\brho=\frac
{\nabla_\bmu P\one(\bmu) e^{L^d \beta P\one(\bmu)}
+\nabla_\bmu P\two(\bmu) e^{L^d \beta P\two(\bmu)}}
{e^{L^d \beta P\one(\bmu)} + e^{L^d \beta P\two(\bmu)}}
\ee
This is of the form (\ref{rhobar}) with the natural correspondences
$\brho\a(\bmu)=\nabla_\bmu P\a(\bmu)$, $x\one=1/(1+r)$,
$x\two=r/(1+r)$ and $\ln r=L^d\beta[P\two(\bmu)-P\one(\bmu)]$. We now
expand near the transition, keeping terms to the same order as in
(\ref{quadr_expansion}). With the abbreviations
$\Delta\brho\cl=\brho\two\cl-\brho\one\cl$ and
$\Delta\bchi=\bchi\two-\bchi\one$ and using that method A imposes
$\bar\brho = n\p\bff\p$, where $\bff\p$ is the normalized parent size
distribution, one has
\bea
n\p \bff\p &=&
\brho\one\cl + \bchi\one\bmu + \frac{r}{1+r}(\Delta\brho\cl +
\Delta\bchi\bmu)
\label{mu_relation}
\\
\ln r &=& L^d\beta(\Delta\brho\cdot\bmu+\bmu\cdot\Delta\bchi\bmu)
\label{r_relation}
\eea
After eliminating $\bmu$, these relations determine the dependence of
$r$ on $n\p$ that we seek. To make progress, we anticipate that
the peak in the second derivative $-(\partial/\partial n\p)^2\ln r$
will occur at a point where $r=O(L^{-d})$. From (\ref{r_relation})
this implies that $\bmu = O(L^{-d}\ln L)$ is also small. Keeping only
leading order terms in the small quantities $r$ and $\bmu$ and using
that $\brho\one\cl = n\p\cl\bff\p$, our previous relations then become
\bea
(n\p-n\p\cl) \bff\p &=& \bchi\one\bmu + r \Delta\brho\cl
\label{mu_relation2}
\\
\ln r &=& L^d\beta\Delta\brho\cl\cdot\bmu
\label{r_relation2}
\eea
Solving the first for $\bmu$ and inserting into the second then gives
\be
\ln r = L^d\beta\Delta\brho\cl\cdot(\bchi\one)^{-1}
[(n\p-n\p\cl) \bff\p - r \Delta\brho\cl] 
\label{relation3}
\ee
To absorb the numerical coefficients and make the parent density
dimensionless we define
\bea
z &=& a r L^d
\label{z_def}
\\
\tilde{n}\p &=& bL^d(n\p-n\p\cl) + \ln(a L^d)
\label{tilde_n_def}
\eea
with
\bea
a &=& \beta\Delta\brho\cl\cdot(\bchi\one)^{-1}\Delta\brho\cl \\
b &=& \beta\Delta\brho\cl\cdot(\bchi\one)^{-1}\bff\p \label{b_def}
\eea
The relation (\ref{relation3}) then becomes just
\be
\tilde{n}\p = z + \ln z
\ee
Differentiating w.r.t.\ $\tilde{n}\p$ gives
$(\partial/\partial\tilde{n}\p)\ln z = (z+1)^{-1}$ and
$-(\partial/\partial\tilde{n}\p)^2\ln z = (z+1)^{-2} z\,
(\partial/\partial\tilde{n}\p)\ln z = z(z+1)^{-3}$. Bearing in mind
that $\ln z$ and $\ln r$ differ only by a constant, we therefore
arrive at a universal large-$L$ scaling form for our second-derivative plot
\begin{equation}
-\left(\frac{\partial}{\partial\tilde{n}\p}\right)^2 \ln r = \frac{z}{(1+z)^3},
\quad
\tilde{n}\p = z+\ln z\:
\label{eq:master}
\end{equation}
which is parameterized by $z$. All dependence on system details is
encoded in the two numerical constants in the definition
(\ref{tilde_n_def}) of the scaled parent density. The curve
(\ref{eq:master}) has its peak at $z=1/2$ so that, from
(\ref{z_def}), $r$ is $O(L^{-d})$ in the region of interest as
anticipated in our derivation. The position of the peak on the
horizontal axis is $\tilde{n}\p=(1/2)-\ln 2$.  The scaling
(\ref{tilde_n_def}) then implies that the cloud point
estimated from the peak position has finite-size corrections of order
$L^{-d}\ln L$, while the peak width and height scale as $L^{-d}$ and
$L^{2d}$, respectively. We will find these scalings, and indeed the
full shape of the master curve (\ref{eq:master}), confirmed in the
simulation data shown below.

\subsection{Method B}
\label{sec:methodB}

While the scaling analysis described above provides a detailed picture
of the finite-size corrections that arise when using method A to
estimate the location of the cloud point, it would clearly be desirable
from a practical point of view to reduce these corrections and ideally
make them exponentially small in system size. For monodisperse phase
coexistence, this is achieved by measuring the densities of the
coexisting phases at the special point where the peaks of $p(n)$ have
equal weights $x\one=x\two=1/2$, i.e.\ $r=1$. At this point the pressure
difference vanishes. In method A, on the other hand, the pressure
difference is $\DP=L^{-d}k_{\rm B}T\ln r\sim L^{-d}\ln L$, and this
causes the relatively large finite-size corrections. This observation
suggests that one should also consider coexisting phases with $r=1$ in
the polydisperse case. Of course, one can then no longer require the
parent density distribution to equal the overall density distribution in
the system, since the latter is an equal mixture of $\ronesig$ and
$\rtwosig$. One has to allow more generally 

\be \parent = (1-\xi)\ronesig + \xi\rtwosig
\label{method_b} 
\ee
where $\xi$ is a parameter to be determined; the cloud point is
estimated as the parent density at which $\xi$ reaches zero.  The
results do, however, have physical meaning also for other values of
$\xi$ in the range $0\leq \xi\leq 1$: they then provide estimates of
properties of the coexisting phases for parent densities $n\p$ within
the coexistence region, with $\xi$ estimating the fractional volume of
phase (2). In simple cases, the cloud point $n\p\cl$ could in fact be
estimated by linearly extrapolating a few measurements of $\xi(n\p)$
to $\xi=0$. In a monodisperse system this would be exact since
$\xi(n\p)$ varies strictly linearly across the coexistence region. In
the polydisperse case, on the other hand, $\xi(n\p)$ is a nonlinear
function and can exhibit very significant curvature near the cloud
point~\cite{e_g_my_review,WilSolFas05,SpeSol03a}. Linear extrapolation
is then unreliable and best avoided in favour of direct determination
of the density $n\p$ where $\xi=0$, as explained above.

We will call this approach ``method B''. In practice, it is implemented
as follows. A parent density is fixed, along with a trial value of
$\xi$. The chemical potentials are then tuned until the density
distributions of the two coexisting phases satisfy (\ref{method_b}). One
measures $r=x\two/x\one$; if this deviates from $r=1$, $\xi$ is adapted
(e.g.\ using a bisection method) and the process is iterated until $r=1$
to numerical accuracy. If a solution with $0<\xi<1$ is found, we are in
the coexistence region. The parent density is then reduced and the
process repeated until $\xi$ drops to zero or no solution with positive
$\xi$ can be found.

It is straightforward to adapt the above scaling analysis to show
that, within method B, finite-size corrections are indeed
exponentially small. The condition (\ref{method_b}) with $\xi=0$
together with $r=1$ gives as the analogues of
(\ref{mu_relation2},\ref{r_relation2})
\bea
(n\p-n\p\cl) \bff\p &=& \bchi\one\bmu
\label{mu_relation_b}
\\
0 &=& L^d\beta\Delta\brho\cdot\bmu
\label{r_relation_b}
\eea
Solving the first and inserting into the second gives
\be
0 = (n\p-n\p\cl) L^d\beta\Delta\brho\cl (\bchi\one)^{-1}\bff\p
\label{relation4}
\ee
which (barring accidental vanishing of the constant factor, equal to
$b$ from (\ref{b_def})) is
satisfied only for $n\p=n\p\cl$. Finite-size corrections therefore
arise only from terms that we had discarded from the outset in our
analysis; these are exponentially small. Note that in a practical
implementation it is important that $\xi$ is determined to high
accuracy; indeed, the analysis above assumes that there is no error in
the value of $\xi$. It is easy to see that, if instead $\xi$
was found only with an accuracy of $O(L^{-d})$, then finite-size
corrections of the same order as in method A arise.

To summarize, method B is algorithmically a little more involved than
method A because it requires for each parent density an ``inner loop''
over $\xi$, but compensates
for this by producing much smaller finite-size corrections. This it
achieves by forcing coexisting phases to have identical
pressures. The parameter $\xi$ has to be introduced to ensure that the
parent distribution is still obtained as an uneven mixture of the two
phases. Note that this is a peculiarity of the polydisperse case: no
such parameter is necessary for monodisperse systems since the properties of
the coexisting phases are the same everywhere within the coexistence
region. In a polydisperse scenario, on the other hand, the coexisting
phases do change~\cite{e_g_my_review}, and so it is important that the
mixing proportions appropriate to the chosen parent density are
maintained.

\section{Application to a polydisperse lattice gas model}

\label{sec:simulations}

\subsection{Model definition}

In order to test the predictions for the finite-size scaling properties
of the two methods detailed above, we have performed a systematic Monte
Carlo simulation study of a novel lattice gas model for a polydisperse
fluid. The choice of a lattice-based rather than a continuum model was
made on the ground of computational tractability: it permits the study
of a larger range of system sizes than is feasible for continuum models.
We do not expect the scaling predictions for the cloud point estimates to be
affected by the presence, or otherwise, of a lattice.

Our polydisperse lattice gas (PLG) model is defined within the grand
canonical ensemble by the hamiltonian:

\begin{equation}
H=-\sum_{ij,\sigma,\sigma^\prime} \sigma \sigma^\prime c_i(\sigma) c_j(\sigma^\prime)-\sum_{i,\sigma}\mu(\sigma)c_i(\sigma).
\label{eq:hamiltonian}
\end{equation}
Here $\sigma$ is the particle ``species'', whose chemical potential is
$\mu(\sigma)$, while $c_i(\sigma)$ is the number
of particles of species $\sigma$ at site $i$, for which we impose a
hard-core constraint such that $\sum_\sigma c_i(\sigma) = 0$ or $1$. The
instantaneous density distribution follows as
$\rho(\sigma)=L^{-d}\sum_ic_i(\sigma)$, with $d=3$ in the simulations
reported below; $i$ runs over the sites of a periodic lattice
$i=1,...,L^{d}$, assumed cubic in this work.
The sum in the first term on the right 
hand side of (\ref{eq:hamiltonian}) similarly runs over all pairs
$i$,$j$ of nearest neighbor sites, as well as
over all combinations of $\sigma$ and $\sigma^\prime$.

For a study of this model under conditions of fixed polydispersity, one
requires knowledge of the chemical potential distribution $\mu(\sigma)$
corresponding to some prescribed form of the ensemble averaged density
distribution $\bar\rho(\sigma)$. In the context of method A, one tunes
$\mu(\sigma)$ such that $\bar\rho(\sigma)=\rho^{(0)}(\sigma)$, while for
method B one simultaneously tunes $\mu(\sigma)$ together with the parameter
$\xi$ to satisfy (\ref{method_b}) {\em and} the equal peak weight
criterion for $p(n)$. Such tuning can be efficiently achieved by a
combination of a non-equilibrium Monte Carlo procedure and histogram
extrapolation techniques, as has been described previously elsewhere
\cite{musig_tuning}. 

We have studied the dilution line properties for a parent distribution
having the Schulz form: 

\begin{equation}
f^{(0)}(\sigma)=\frac{1}{Z!}\left( \frac{Z+1}{\bar{\sigma}} \right)^{Z+1} \sigma^{Z} \exp\left[ - 
\left( \frac{Z+1}{\bar{\sigma}} \right) \sigma\right].
\label{schulz}
\end{equation}
The mean value of the distribution $\bar{\sigma}$ defines the unit length, while
the parameter $Z$ controls the width of the distribution. We fixed the
latter to be $Z=50$, resulting in a dimensionless degree of
polydispersity 
\begin{equation}
\delta\equiv \frac{\sqrt{\overline{(\sigma-\bar{\sigma})^2}}}{\bar\sigma}=\frac{1}{\sqrt{Z+1}}\simeq14\%
\label{eq:poly}
\end{equation}
Lower and upper cutoffs were imposed on the distribution 
at $\sigma=0.5$ and $1.4$ respectively, and the distribution was normalized accordingly. 


\subsection{Simulation results}

A determination of the critical point parameters for the PLG model using
well-established finite-size scaling methods \cite{WILDING95} found the
critical temperature to be located at
$(n_{c}^{(0)},T_{c})=(0.521(1),1.171(1))$ in reduced units. This
is to be compared with the critical parameters of the monodisperse
(Ising) lattice gas $(0.5,1.127955)$ \cite{LUIJTEN}. Thus the inclusion
of polydispersity of the form (\ref{eq:hamiltonian}) is seen to raise
both the critical temperature and the critical density. Moreover it
splits the liquid-gas binodal into well separated cloud and shadow
curves  (cf.\ fig.~\ref{fig:phasediagram}) in a manner similar to that
occurring in continuum fluid models for which polydispersity affects the
interparticle interaction strength as well as its range
\cite{WilSolFas05,WilFasSol04}. As a consequence, the critical point
lies below the maximum of the cloud curve and phase coexistence can be
observed even at $T=T_c$, provided that $n^{(0)}<n_c^{(0)}$. 

In view of this we have adopted the critical temperature as a convenient
reference point and have studied coexistence along the dilution line
$\rho^{(0)}(\sigma)=n^{(0)}f^{(0)}(\sigma)$ for fixed $T=T_c$. This
involved performing a series of MC runs starting from the critical
density and reducing $n^{(0)}$ in a stepwise fashion. Multiple histogram reweighting
techniques were employed to estimate the chemical potential distribution
at each successive step, while use of multicanonical preweighting
techniques \cite{BergNeuh92} ensured that both coexisting phases were
efficiently sampled in the course of each simulation run (see
ref.~\cite{WilFasSol04} for a fuller account of this procedure). 

\begin{figure}[h]
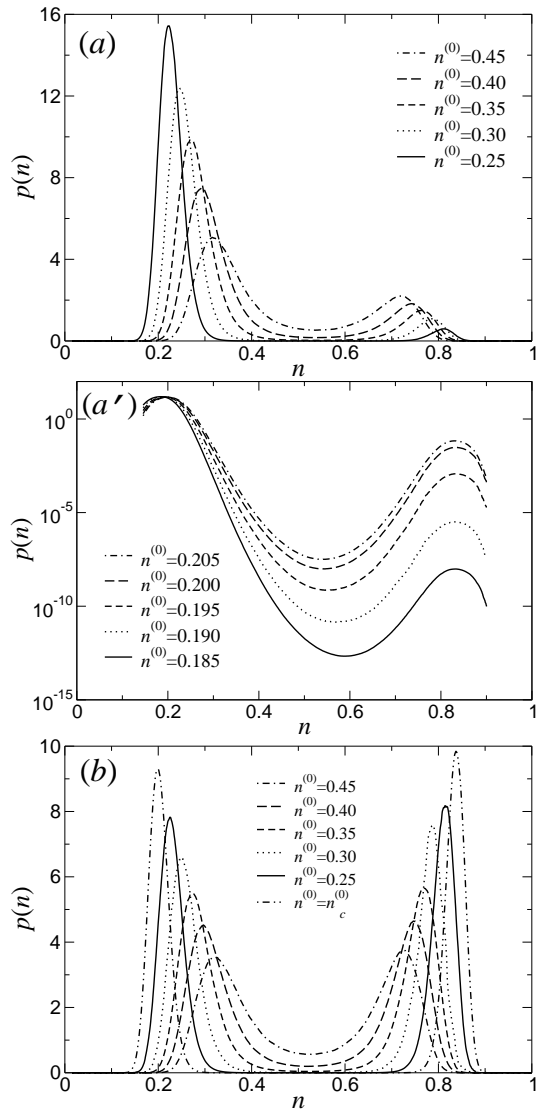

\includegraphics[width=7cm,clip=true]{pnop1a.eps}
\includegraphics[width=7cm,clip=true]{pnop1b.eps}
\includegraphics[width=7cm,clip=true]{pnops2.eps}
\caption{${\bf (a)}$ Estimates of the form of $p(n)$ for a
selection of values of $n^{(0)}$ within the liquid-vapor coexistence
region as obtained from the GCE simulation data obtained according to
method A. The associated estimates of $r$ are, in order of decreasing
$n^{(0)}$: $r=0.4483, 0.2682, 0.1643, 0.0901,  0.0379$. ${\bf (a')}$ The distributions for
a selection of small values of
$n^{(0)}$ displayed on a logarithmic scale.  ${\bf (b)}$ The
corresponding estimates of $p(n)$ obtained according to method B (for
which the peak weight ratio is constrained to unity). The associated estimates of $r_\xi$ are, in order of decreasing
$n^{(0)}$: $r_\xi=0.4158, 0.2411, 0.1477, 0.0822,  0.0335$. All the plots
refer to system size $L=15$.}
\label{fig:pn_methods}
\end{figure}

The dilution line was scanned in this manner for lattices of sizes
$L=10,~12,~15,~18,~21$ using both methods A and B. The GCE simulations
directly yield the form of $p(n)$ corresponding to either
case. For method A, one observes (figs.~\ref{fig:pn_methods}(a) and
\ref{fig:pn_methods}($a^\prime$)) that as $n^{(0)}$ is reduced from its
critical point value, the peaks in $p(n)$ separate, while the valley
between them deepens. This is accompanied by a gradual transfer of
weight from the liquid to the gas peak. We extract 
the ratio of the peak areas at a given $n^{(0)}$ from
$p(n)$ via

\begin{equation}
r(n^{(0)})=\frac{\int_{n>n^{*}}p(n)dn}{\int_{n<n^{*}}p(n)dn},
\label{r_determination}
\end{equation}
Here $n^{*}$ is a convenient threshold density intermediate between vapor
and liquid densities, which we take to be the location of the minimum
in $p(n)$.

The form of $p(n)$ obtained using method B is shown in
fig.~\ref{fig:pn_methods}(b) for a selection of values of $n^{(0)}$.
Here, by virtue of the appropriate tuning of the parameter $\xi$, the
vapor and the liquid peaks maintain equal weights throughout the
coexistence region. The estimate of the ratio of fractional phase
volumes is obtained simply as $r_\xi=\xi/(1-\xi)$. We use the
subscript $\xi$ here to distinguish this quantity from the probability
mass ratio $r$ as
extracted from (\ref{r_determination}), the latter
being more directly related to the pressure
difference between the phases as explained above.

\begin{figure}[h]
\includegraphics[width=8.5cm,clip=true]{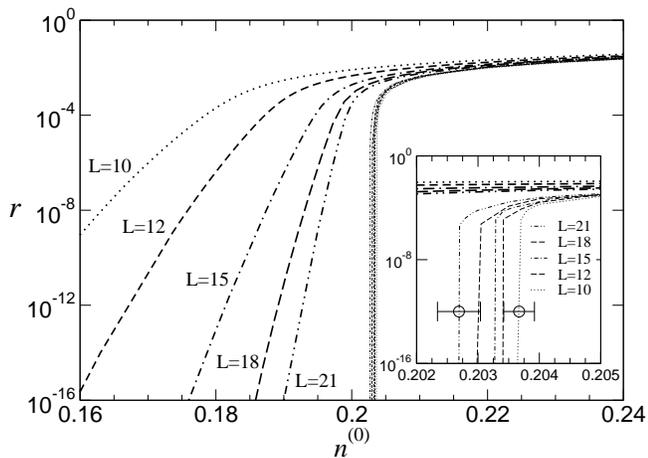}
\caption{The value of $r$ as estimated from method A (labelled
thick curves) and of $r_\xi$ from method B (unlabelled, thin curves);
both are plotted against parent density $n\p$. The
inset shows in greater detail the results for method B in the vicinity
of the cloud point density. An estimate of the
statistical errors of the procedure were obtained via a block analysis.
The errors for the smallest and largest system size are indicative of
the statistical uncertainty of our results. 
}
\label{fig:ratios}
\end{figure}

In fig.~\ref{fig:ratios}, we plot the dependence of $r$ on parent
density $n^{(0)}$ as obtained from method A for the various
system sizes we have studied, alongside the results for $r_\xi$ from
method B. The curves for method A (thick lines)
show a strong $L$-dependence within the metastable coexistence
region which borders the cloud point at small $n^{(0)}$.
As $n^{(0)}$ is increased, however, and the 
cloud point is approached, the curves cross over to their
$L$-independent limit values in the coexistence region.

\begin{figure}[h]
\includegraphics[width=8.5cm,clip=true]{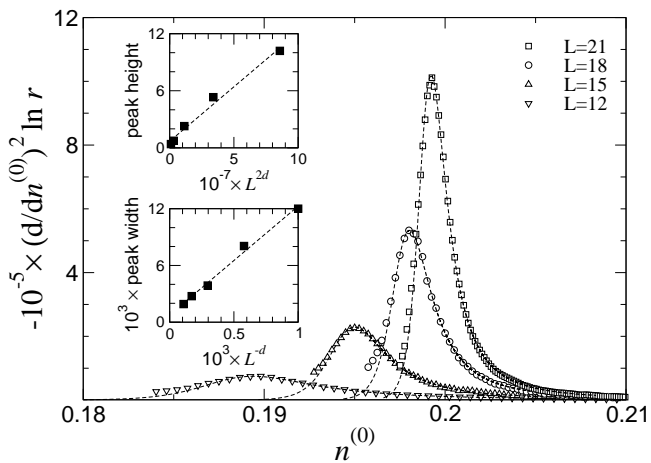} 
\caption{Points show the estimated second derivative $-(\partial /
\partial n^{(0)})^{2} \ln r$ obtained within method
A. The solid lines superimposed on these data points are scaled fits
of the form (\protect\ref{eq:master}). Insets show the
$L$-dependence of the peak width and height plotted in terms of the
predicted scaling variable and accompanied by a least squares fit.}
\label{fig:masterfits} 
\end{figure}

Looking in more detail at the finite-size scaling properties of the curves for
$r(n^{(0)})$ from method A, the analysis of sec.~\ref{sec:methodA}
shows that the cloud point region is associated with a peak in the second
derivative $-(\partial / \partial n^{(0)})^{2} \ln r$. In
Fig.~\ref{fig:masterfits} we plot this quantity for the various systems
sizes studied. Superimposed on each plot is a suitably scaled form of
the predicted universal master curve, (\ref{eq:master}). The
non-universal scale factors for the height and the width of the master
curves implicit in the fitting shown are plotted in the insets against
the predicted scaling variables $L^{2d}$ and $L^{-d}$ respectively (cf.
sec.~\ref{sec:methodA}).  Clearly there is good agreement with the
predictions for both the general shape of the peak and the scaling of
its width and height. Some discrepancies between the observed and
measured peak shape are apparent on the low density side, well away from
the peak maximum, particularly for the smaller system sizes. These
are presumably attributable to the breakdown of the validity of
the linear expansion in the density difference used in the derivation of
(\ref{eq:master}).

The estimates of $r_\xi(n^{(0)})$ deriving from application of method B,
as shown in Fig.~\ref{fig:ratios}, exhibit a qualitatively different
behavior. As the density is increased from the cloud point,
one observes a rapid, near-vertical increase in
$\ln r_\xi$
from a large negative value (where $r_\xi$ is essentially zero) to
values of $O(1)$. There is only a
weak dependence of this behaviour on the system size (see inset of
fig.~\ref{fig:ratios}). Within this method the cloud point density can
thus be directly read off as the lowest density at which an
equal-peak-weight solution exists to the numerical procedure described
in Sec.~\ref{sec:methodB}. We note that, as expected, $r$ from method
A and $r_\xi$ from method B converge to similar values for parent
densities well within the coexistence region.

\begin{figure}[h]
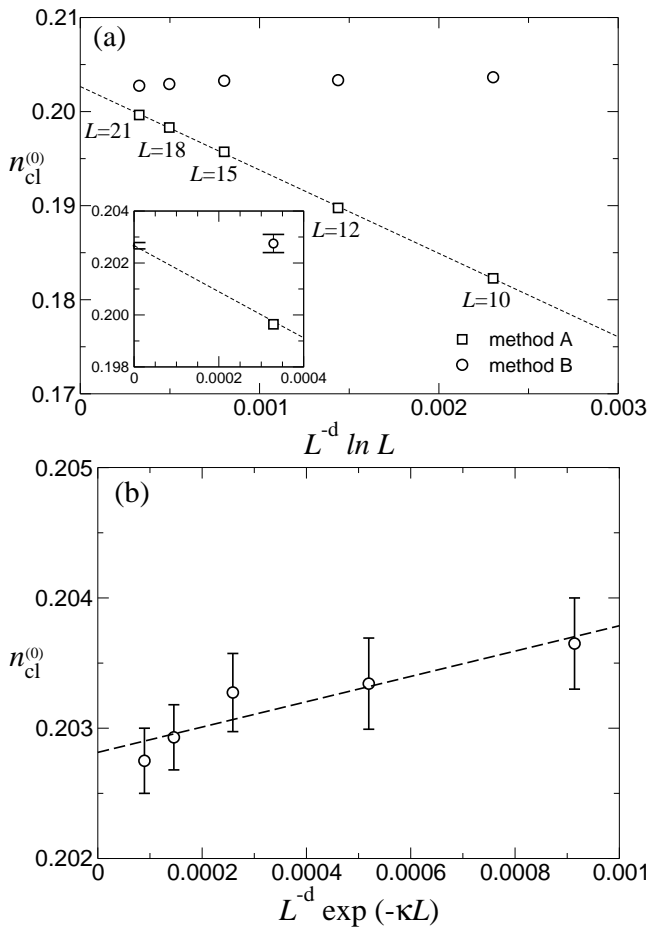

\includegraphics[width=8.5cm,clip=true]{fss_vlogv_new.eps}
\includegraphics[width=8.5cm,clip=true]{exp_scaling.eps}
\caption{Finite-size scaling of the cloud point estimates deriving from
methods A and B. {\bf (a)} The data for both methods are plotted against
the predicted scaling variable for method A. The dashed line is a least
squares fit to the estimates from method A and the inset shows a
magnified view of the region of large $L$. {\bf (b)} The estimates
deriving from method B, plotted against $L^{-d}\exp(-\kappa L)$ with $\kappa=8.988\times 10^{-3}$. Except where error bars are shown, statistical uncertainties are smaller than
the symbol sizes.}
\label{fig:fss} 

\end{figure}

Fig.~\ref{fig:fss} compares for the two methods the finite-size
behaviour of the cloud point density estimates. In the case of method A,
the estimate for a given $L$ derives from the position of the peak in
the second derivative plot (fig.~\ref{fig:masterfits}).
Fig.~\ref{fig:fss} confirms that (as predicted) these estimates deviate
from their limiting value by a correction O$(L^{-d}\ln L)$. A
least-square fit yields $n\cl^{(0)}=0.20266(12)$ as the best estimate
of the cloud point density (see also inset). In the case of method B,
the cloud point is estimated as the lowest value of $n^{(0)}$ for which
an equal-peak-weight solution for $p(n)$ can be found. The finite-size
corrections to this estimate are expected to be exponentially small in
the system size and our results (fig.~\ref{fig:fss}(b)) are consistent
with this, yielding a cloud point estimate $n\cl^{(0)}=0.2028(1)$.
Indeed the corrections appear to be so small that even for the smallest
system size studied ($L=10$) the estimate of the cloud point density
obtained using method B deviates from the limiting value by just
$0.5\%$. This compares with a relative deviation of approximately $10\%$
for method A.

\section{Conclusions}
\label{sec:concs}

We summarize. We have presented and tested two distinct finite-size
scaling strategies for estimating coexistence properties and cloud
points in polydisperse fluids within a grand canonical ensemble
framework. Both are, of course, also directly applicable to multicomponent 
mixtures of many discrete species. The first approach, ``method A''
takes the natural step of
constraining the ensemble average of the density distribution
$\rho(\sigma)$ to match the prescribed parent form. However, in
order to satisfy the lever rule, this necessarily leads to unequal peak
weights in the order parameter distribution function $p(n)$. For systems
of finite size, this translates to a finite pressure difference between
the coexisting phases, which in turn engenders finite-size corrections
to estimates of the cloud point density which are {\em powers} of the
system size. 

The second approach, ``method B'' strictly imposes equal peak weights
for $p(n)$, and hence pressure equality in systems of finite size.  The
coexistence properties of the desired parent are obtained by
appropriately weighting the relative contributions of the coexisting
phases to the overall density distribution in such a way that the lever
rule is satisfied. We have detailed how to determine this weight factor
operationally and shown that method B has finite-size corrections to
the coexistence properties which are {\em exponentially small} in the
system size. As such, and notwithstanding the need to determine the
phase weighting factor $\xi$, method B is clearly superior to method A.
It can be regarded as a generalization to multicomponent mixtures of the
well known equal peak weight criterion \cite{BorKot92} developed in the
context of monodisperse systems by Borgs and coworkers.

\vspace*{5mm} 
\acknowledgments

This work was supported by the EPSRC, grant number GR/S59208/01. MM
acknowledges the support of the Volkswagen foundation.


\vskip2cm
\begin{thebibliography}{99}

\bibitem{SALACUSE} J.J.~Salacuse and G.~Stell, J. Chem. Phys. {\bf 77},
3714 (1982).

\bibitem{LARSON99} R.G.~Larson, {\em The Structure and Rheology of
Complex fluids} (Oxford University Press, New York, 1999).

\bibitem{CHAIKIN00} P.~Chaikin in {\em Soft and Fragile Matter}, M.E.
Cates and M.R. Evans (eds.), IOP publishing, London, 2000.

\bibitem{e_g_my_review} P.~Sollich, J. Phys. Condens. Matter {\bf 14}, R79 (2002). 

\bibitem{NOTE0}  For each parent density within the cloud curve there is a
unique coexistence curve, not shown in fig.~\ref{fig:phasediagram}; see ref.~\protect\cite{e_g_my_review}.

\bibitem{WILDING95} N.~B.~Wilding, Phys. Rev. E {\bf 52}, 602 (1995).

\bibitem{WilSolFas05} N.B.~Wilding, P.~Sollich and M.~Fasolo, Phys. Rev. Lett. {\bf 95}, 155701 (2005).

\bibitem{musig_tuning} N.B.~Wilding, J. Chem. Phys. {\bf 119}, 12163 (2003); N.B.~Wilding and P.~Sollich,
J. Chem. Phys. {\bf 116}, 7116 (2002).

\bibitem{BorKot92} C.~Borgs and W.~Janke, Phys. Rev. Lett. {\bf 68}, 1738 (1992); C.~Borgs and R.~Kotecky,
Phys. Rev. Lett. {\bf 68}, 1734 (1992).

\bibitem{SpeSol03a} A.~Speranza and P.~Sollich, J.\ Chem.\ Phys. {\bf
118}, 5213 (2003).


\bibitem{LUIJTEN} H.W.J.~Bloete, E.~Luijten and J.R.~Heringa, J. Phys. A 28, 6289 (1995).

\bibitem{WilFasSol04} N.B.~Wilding, M.~Fasolo and P.~Sollich, J. Chem. Phys. {\bf 121}, 6887 (2004). 

\bibitem{BergNeuh92} B.A.~Berg and T.~Neuhaus, Phys. Rev. Lett. {\bf 68}, 9 (1992).

\end{thebibliography}
\end{document}